# Design of a Plasmonic Absorption Electro-optical Modulator Based on n-doped Silicon and Barium Titanate

Purya Es'haghi[a*], Ali Barkhordari[b] and Abolfazl Safaei Bezgabadi[a]

[a]*Photonics Department, Graduate University of Advanced Technology, Kerman, Iran;*
[b]*Department of plasma engineering, Graduate University of Advanced Technology, Kerman, Iran*

*Corresponding Author: puryae@gmail.com

**Abstract:**
In this paper, a numerical solution for a plasmonic absorption modulator with a six-layer structure consisting of an air superstrate, a gold layer, a barium titanate layer, a n type silicon layer, a gold layer and a thin-shell nanolattice aluminium oxide substrate is presented. Regarding the suggested structure, parameters related to the absorption modulation are investigated at different thicknesses. Here, the Pockels effect and the free carrier dispersion effect are considered simultaneously. The dispersion relation of this structure is analytically obtained and numerically solved by the Nelder-Mead method. The maximum calculated figure of merit is 12.79. Furthermore, according to our results, it is understood that this modulator has a high ability to be utilized in optical communication systems. Also, it could be integrated to the microelectronic systems and it is compatible with CMOS technology.

## 1. Introduction

According to Moore's law, the number of transistors in a fixed area of dense integrated circuit doubles almost every two years and consequently electronic devices are going to be more compact. Because of the diffraction limit, optical waveguides do not have compaction capability below the half of propagation wavelength. Hence, they can't be integrated with electronic devices. This is a defect in waveguide-based systems. Recently, scientists have turned to plasmonic waveguides in order to overcome the diffraction limit [1-3].

It is important to use the plasmonic waveguide-based modulators and switches As these modulators and switches have footprints on the order of several $\mu$ m$^2$, they have the capability to cointegrate with electronic systems [4,5]. These types of modulators and switches have many advantages, such as high bandwidth [6], low power consumption [7], and ultra-compact size [8]. Also, these modulators and switches mainly employ the thermal optics effect [9-11], the free carrier dispersion effect [12,13], the Pockels effect [14-16], the phase transition effect [17-19] and the electrochemical metallization effect [20,21]. Owing to the usage of different types of waveguides based on surface plasmon polaritons, these devices have a strong interaction between propagating optical signals and active materials [22-27], which is necessary to produce optical switches well.

As silicon-based Metal-Insulator-Metal waveguides can be created with a volumetric size of a cubic wavelength they are suitable for planar integration. Hence, they are compatible with CMOS technology and are able to integrate with microelectronics devices. The modulators based on these waveguides need a low operation voltage. These offer a large modulation depth and furthermore provide a switching speed as fast as several gigahertz [28].

Here, a silicon-based plasmonic modulator is presented which it has an Air/Au/BaTiO$_3$/n-Si/Au/thin-shell nanolattice Al$_2$O$_3$ configuration. This structure is studied as an intensity modulator at the 1.55μm wavelength. The dispersion relation for this structure is obtained and solved for 0 and 10V applied voltages in order to analyze its properties. Therefore, the Poisson equation is solved to apply an electric field to the voltage-dependent variation of the refractive index. It should be noted that the free carrier dispersion effect and the Pockels effect are simultaneously considered in this device.

## 2. Theory

### 2.1. Modulator Structure

As it is understood, the presented structure is a slab waveguide. This waveguide consists of two layers of n-type Silicon and barium titanate which are embedded between two layers of Au. The substrate of this structure is a thin-shell nanolattice aluminium oxide (TSNAO) with near unity refractive index. The above region of the structure is filled with the air. The BaTiO$_3$ is a birefringent crystal that matches its principal-axis system to the axes of the considered coordinate system. Figure 1 shows the structure of the investigated modulator. The refractive indices of each layer, which form the modulator, can be given by:

$$n_{TSNAO} = 1.025 \quad (1)$$

$$n_{Au}^2 = 1.53 - \frac{1}{145^2\left(\frac{1}{\lambda^2} + \frac{i}{17000\lambda}\right)} + \frac{0.94}{468}\left[\frac{e^{-i\pi/4}}{\left(\frac{1}{468} - \frac{1}{\lambda} - \frac{i}{2300}\right)} + \frac{e^{-i\pi/4}}{\left(\frac{1}{468} + \frac{1}{\lambda} + \frac{i}{2300}\right)}\right] + \frac{1.36}{331}\left[\frac{e^{-i\pi/4}}{\left(\frac{1}{331} - \frac{1}{\lambda} - \frac{i}{940}\right)} + \frac{e^{-i\pi/4}}{\left(\frac{1}{331} + \frac{1}{\lambda} + \frac{i}{940}\right)}\right] \quad (2)$$

$$\varepsilon_{n-Si} = \varepsilon_\infty - \frac{\omega_p^2}{[\omega^2(1+\frac{1}{\omega\tau})]}, \quad \omega_p^2 = \frac{N_i e^2}{\varepsilon_0 m^*}, \quad \tau = \frac{m^*\mu}{e} \quad (3)$$

where $\mu = 50 \, cm^2/Vs$, $m_e = 9.1 \times 10^{-31} kg$, $m^* = 0.272 m_e$, $\varepsilon_\infty = 11.7$, $e = 1.602 \times 10^{-19} C$ and $\varepsilon_0 = 8.85 \times 10^{-12} \, C^2/Nm^2$ for substrate Au layer and n-Si layer, respectively [29-31]. Also, the ordinary and extraordinary refractive indexes of the BaTiO$_3$ can be written as [32]:

$$n_o^2 = 3.05840 + \frac{2.27326 \lambda^2}{\lambda^2 - 0.07409} - 0.02428 \lambda^2 \quad (4)$$

$$n_e^2 = 3.02479 + \frac{2.14062 \lambda^2}{\lambda^2 - 0.067007} - 0.02169 \lambda^2 \quad (5)$$

In equations (1) to (5), n, $\lambda$, $\varepsilon_\infty$, $\omega_p$, $\tau$, $\mu$, $m_e$, $m^*$, $\varepsilon_0$, e and N are the refractive index, the wavelength, the background permittivity, the plasma frequency, the relaxation time, the electron mobility, the electron mass, the electron effective mass, the vacuum permittivity, the electron charge and the free carrier density, respectively. The free carrier density is approximately equal to the doping concentration in the heavily doped n-Si. In this case the doping concentration is equal to $10^{21} cm^{-3}$ [33].

Surface Plasmon-Polaritons (SPPs) are propagated between two gold layers. By applying a voltage between the two layers, a static electric field is created inside the dielectric layers. This field changes the refractive index of $BaTiO_3$, it also causes a shift in the real and imaginary parts of the n-Si refractive index. These variations also lead to alter the intensity of surface plasmon polaritons. To investigate these evolutions, it's necessary to obtain the dispersion relation.

## 2.2. Dispersion relation

In order to determine the dispersion relation for the structure in the on and off states, the TM mode solutions are used in the wave equation. By assuming that the time dependence part of the electromagnetic field is $e^{-i\omega t}$, the wave equation for the magnetic field component can be described by [34]:

$$\vec{\nabla} \times (\overleftrightarrow{\varepsilon}^{-1} \vec{\nabla} \times \vec{H}) - \omega^2 \mu_0 \vec{H} = 0 \qquad (6)$$

As it is required to stimulate the TM modes for SPP modes, the wave equation is given by [35]:

$$\frac{\partial^2 H_y}{\partial z^2} + \left(k_0^2 \varepsilon_x - \beta^2 \frac{\varepsilon_x}{\varepsilon_z}\right) H_y = 0 \qquad (7)$$

The components of the electric field for the TM modes are:

$$E_x = \frac{-i}{\omega \varepsilon_0 \varepsilon_x} \frac{\partial H_y}{\partial z} \qquad (8)$$

$$E_z = \frac{-\beta}{\omega \varepsilon_0 \varepsilon_z} H_y \qquad (9)$$

The n-Si layer is divided into 10 parts for the more accurate study of electrostatic and electrodynamic field effects. By assuming that the time and the x-coordinate dependent parts are in $e^{-i(\omega t - \beta x)}$ form, the solutions which satisfy equation (7) are:

$$H_y = \begin{cases} A_1 e^{ik_1(z-z_1)} & -\infty < z < z_1 \\ A_2 e^{ik_2(z-z_2)} + A_3 e^{-ik_2(z-z_2)} & z_1 < z < z_2 \\ A_4 e^{ik_3(z-z_3)} + A_5 e^{-ik_3(z-z_3)} & z_2 < z < z_3 \\ A_6 e^{ik_4(z-z_4)} + A_7 e^{-ik_4(z-z_4)} & z_3 < z < z_4 \\ \vdots & \vdots \\ A_{22} e^{ik_{12}(z-z_{12})} + A_{23} e^{-ik_{12}(z-z_{12})} & z_{12} < z < z_{13} \\ A_{24} e^{ik_{13}(z-z_{13})} + A_{25} e^{-ik_{13}(z-z_{13})} & z_{13} < z < z_{14} \\ A_{26} e^{ik_{14}(z-z_{14})} & z_{14} < z < +\infty \end{cases} \qquad (10)$$

where $z_1 = 0$, $z_2 = h_{Au}$, $z_3 = h_{Au} + \frac{h_{n-Si}}{10}$, $z_4 = h_{Au} + \frac{2h_{n-Si}}{10}$, ... , $z_{12} = h_{Au} + \frac{10h_{n-Si}}{10}$, $z_{13} = h_{Au} + h_{n-Si} + h_{BaTio_3}$, , $z_{14} = h_{Au} + h_{n-Si} + h_{BaTio_3}$. The components of the electric field of the TM mode can be obtained by inserting $H_y$, relation (10), into equations (8) and (9). In relation (10), k is the transverse propagation constant, which for all layers except the $BaTiO_3$ layer, is equal to:

$$k_m = \sqrt{\varepsilon_m k_0^2 - \beta^2} \qquad (11)$$

due to $\varepsilon_x = \varepsilon_z$, and for the $BaTiO_3$ layer is equal to [35]:

$$k_{13} = \sqrt{\varepsilon_x k_0^2 - \beta^2 \frac{\varepsilon_x}{\varepsilon_z}} \qquad (12)$$

where $\beta$ is the propagation constant of SPP mode and $\varepsilon_x$ and $\varepsilon_z$ are the permittivities of the anisotropic layer in the x and z directions respectively also $k_0$ is the wavenumber in the free space. By applying the boundary conditions to $H_y$ and $E_z$, one can obtain

$$M_{26\times26}(\lambda, \beta) \cdot A_{26\times1} = 0 \quad (13)$$

where $\lambda$ is the vacuum wavelength and $A_{26\times1}$ is a column matrix that describes the constants in front of the exponential functions in relation (10).

In order to achieve the dispersion relation, the determinant of $M_{26\times26}$ should be zero [36]. The propagation constant, which it could be a complex number, is acquired by solving the dispersion relation per wavelength.

## 2.3. Solving method

Here, the Nelder-Mead method [37] is applied to solve the dispersion relation as this is an optimization method and it is not needed to take differentiation for finding the value of the parameters for minimizing the absolute value of the dispersion relation. The standard error to stop the numerical calculation is $\epsilon = 2.2204 \times 10^{-16}$. The absorption coefficient and the effective propagation length are respectively calculated as:

$$\alpha = 2\text{Im}[\beta] \quad (14)$$

and

$$L_e = \frac{1}{\alpha} \quad (15)$$

The 1dB on-off length can be defined as:

$$L_{1dB} = \frac{1}{(\log_{10} e)\Delta\alpha} \quad (16)$$

where $\Delta\alpha = |\alpha_{off} - \alpha_{on}|$ is the difference of the absorption coefficients in which $\alpha_{off}$ and $\alpha_{on}$ are the absorption coefficients of the off state and the on state, respectively. The extinction ratio,

$$E_R = (\log_{10} e)\Delta\alpha L, \quad (17)$$

is equal to 1dB in an arbitrary chosen DC voltage where L is the device length. At last, it is worthy to illustrate a parameter for comparing the efficiency of the presented absorption plasmonic modulator based on the above-defined quantities called the figure of merit (FoM) [36,38]:

$$\text{FoM} = \frac{L_e}{L_{1dB}} = (\log_{10} e)\frac{|\alpha_{off} - \alpha_{on}|}{\alpha_{state}} \quad (18)$$

where, $\alpha_{state}$, is the residual absorption coefficient [38].

## 2.4. Applied static electric field

In general, the nonlinear Poisson equation is given by [39]:

$$\vec{\nabla} \cdot \overleftrightarrow{\varepsilon_s}\vec{\nabla}\varphi = -\frac{\rho}{\varepsilon_0} \quad (19)$$

for applying the static electric field in an anisotropic media. The charge density is zero ($\rho = 0$) for the BaTiO$_3$ layer. The nonlinear Poisson equation of this case is solved in one dimension. The electric potential at the interfaces can be acquired by the following boundary conditions:

$$\frac{U}{d} = \varepsilon_{s,z} \frac{d\varphi}{dz}\bigg|_{z\to+(h_{Au}+h_{n-Si})} = \varepsilon_{s,n-Si} \frac{d\varphi}{dz}\bigg|_{z\to-(h_{Au}+h_{n-Si})} \quad (20)$$

$$\varphi(z) = 0 \quad \text{for} \quad z \to +h_{Au} \quad (21)$$

and

$$\varphi(z) = U \quad \text{for} \quad z = h_{Au} + h_{n-Si} + h_{BaTiO_3} \quad (22)$$

where $\phi$ is the electrical potential, $\varepsilon_{s,z}$ is the relative static permittivity of the BaTiO3 in z-direction and $\varepsilon_{s,n-Si}$ is the relative static permittivity of the n-Si.

For the n-Si layer, via exploiting the Thomas- Fermi screening theory [36], the Poisson equation can be rewritten as:

$$\nabla^2 \varphi = \frac{e(N_i(z) - N_0)}{\varepsilon_0 \varepsilon_{s,n-Si}} \quad (23)$$

where

$$N_i(z) = \frac{1}{3\pi^2}\left(\frac{8\pi^2 m^*}{h^2}\right)(E_f + e\varphi(z)) \quad (24)$$

$$\varepsilon_{s,n-Si} = 11.688 + \frac{1.635 \times 10^{-19} N_D}{1 + 1.172 \times 10^{-21} N_D} \quad (25)$$

$$\varepsilon_{s,z} = 135 \quad (29)$$

The Fermi energy, $E_f$, in equation (24) is defined as:

$$E_f = \left(\frac{h^2}{8\pi^2}\right)[3\pi^2 N_0]^{2/3} \quad (26)$$

where $N_0$ is the bulk free carrier density of the n-Si, $N_D$ is the concentration of donors in the n-Si, $m^*$ is the electron effective mass of the n-Si, $E_f$ is the Fermi energy of the n-Si, h is the Planck constant, $N_i$ is the free carrier density of the n-Si [36,40].

The finite difference method with the meshing number of 6000 is applied for solving equation (23) for the n-Si layer [41]. The effects of applying voltage on the refractive index of the n-Si layer can be studied by solving the Poisson equation in this layer where $N_i$ is obtained in 10 parts of the n-Si layer and thus by substituting $N_i$ in relation (3)

For the BaTiO3 layer, the boundary conditions (20) and (22) are used to solve equation (27). Accordingly, the obtained static electric field induces an alteration in the ordinary and extraordinary refractive index which can be characterized by:

$$\frac{d^2 \varphi}{dz^2} = 0 \quad (27)$$

$$\begin{bmatrix} \Delta(\frac{1}{n^2})_1 \\ \Delta(\frac{1}{n^2})_2 \\ \Delta(\frac{1}{n^2})_3 \\ \Delta(\frac{1}{n^2})_4 \\ \Delta(\frac{1}{n^2})_5 \\ \Delta(\frac{1}{n^2})_6 \end{bmatrix} = \begin{bmatrix} r_{11} & r_{12} & r_{13} \\ r_{21} & r_{22} & r_{23} \\ r_{31} & r_{32} & r_{33} \\ r_{41} & r_{42} & r_{43} \\ r_{51} & r_{52} & r_{53} \\ r_{61} & r_{62} & r_{63} \end{bmatrix} \begin{bmatrix} E_x \\ E_y \\ E_z \end{bmatrix} = \overleftrightarrow{r} \cdot \vec{E} \quad (28)$$

and

$$\overleftrightarrow{\eta} = [\frac{1}{n^2}] = \begin{bmatrix} \frac{1}{n_x^2} + \Delta(\frac{1}{n^2})_1 & \Delta(\frac{1}{n^2})_6 & \Delta(\frac{1}{n^2})_5 \\ \Delta(\frac{1}{n^2})_6 & \frac{1}{n_y^2} + \Delta(\frac{1}{n^2})_2 & \Delta(\frac{1}{n^2})_4 \\ \Delta(\frac{1}{n^2})_5 & \Delta(\frac{1}{n^2})_4 & \frac{1}{n_z^2} + \Delta(\frac{1}{n^2})_3 \end{bmatrix} \quad (29)$$

where $\overleftrightarrow{r}$ is the linear electro-optical coefficients matrix (Pockels) described in the standard crystallographic coordinate system that the z-direction indicates the crystal optical axis[42]. Here, $r_{13} = r_{23} = 19.5[\frac{pm}{V}]$, $r_{33} = 97[\frac{pm}{V}]$, $r_{42} = r_{51} = 1640[\frac{pm}{V}]$ and the other coefficients are zeroes [43]. $\overleftrightarrow{\eta}$ is the electro-optically perturbed impermeability tensor where $\Delta(\frac{1}{n^2})_m$ is the electro-optically induced variation of the impermeability tensor element, and $n_x$, $n_y$, and $n_z$ are the refractive indices in x, y, and z-directions, respectively. The resulted refractive indices by applying the voltage are used for solving the dispersion relation in the off state.

## 3. Results and Analysis

### 3.2. Absorption Modulation

Regard to the variation of the imaginary part of the propagation constant of the proposed structure with the applied voltage, it can be utilized as an absorption modulator. The absorption coefficient of the modulator for the on and off states and its dependence to the thickness of the layer are shown in Fig. 2. According to fig. 2, it is understood that the absorption coefficient increases by applying the voltage and it is switched from the on state to the off state. By increasing the thickness of the n-Si layer the absorption coefficient of the on state has small changes with respect to the off state. Fig. 2 also indicates that the absorption coefficient of the on state increase by decreasing the thickness of the n-Si layer but this is not true for the off state. The maximum value of the absorption coefficient for the off state occurs at the 12nm thickness of the n-Si layer while it happens at the 10nm thickness for the on state.

The absorption coefficient of the on and off states increase by decreasing the thickness of the $BaTiO_3$ layer. In addition, by comparing the figs. 2(a), 2(b) and 2(c), it is found that the absorption coefficients of the both states decrease by increasing the thickness of the gold layer.

As high value of the figure of merit (FoM) is appreciated for the absorption modulators, it is necessary to increase the effective propagation length and decrease the 1dB on-off length. Fig. 3 shows the effective propagation length for different thicknesses of the n-Si, the $BaTiO_3$ and the gold layers in the on state. It is indicated that the effective propagation length gets larger by increasing the thicknesses of the layers. In fig. 4, it is shown that the thickness of the gold layer does not have a significant effect on the 1dB on-off length. But increasing the thickness of the $BaTiO_3$ layer change the 1dB on-off length inversely. Also, the minimum value of the 1dB on-off length occurred at a thickness of 15nm for the n-Si layer.

As mentioned earlier, the FoM parameter of the modulators depends on the effective propagation length and the 1dB on-off length. Here, one can discuss the FoM parameter of the presented modulator by the obtained results of fig. 4. The FoM of our modulator for different thicknesses is presented in fig. 5. Increasing the thicknesses of the gold and the $BaTiO_3$ layers result in growing the FoM parameter. The optimized configuration for the modulator is specified by the maximum value of the FoM parameter. The maximum value of the FoM is 12.79 which is obtained at an Air(∞)/Au(50nm)/$BaTiO_3$(30nm)/n-Si(13nm)/Au(50nm)/TSNAO(∞) multilayer configuration.

This value of FoM is much larger than the FoM of the previously reported device in ref. [36] which it expresses that our modulator has higher performance than the presented modulator in that reference as the effective propagation length and the 1dB on-off length have significant contributions on the performance of modulators. The origin of this value of FoM for our modulator is related to the high value of refractive index of $BaTiO_3$ which has an impact on transferring a large amount of energy to the n-Si layer. So, the refractive index of n-Si layer and consequently the absorption coefficient of the configuration change a lot. In addition, as the static permittivity of $BaTiO_3$ is high the electrical power for changing the refractive index of active materials will be reduced. Fig. 6(a) and fig. 6(b) show the propagation constant and the absorption coefficient for the optimized configuration in the on and the off states, respectively. The electric and the magnetic fields, as well as the time-averaged Poynting's vector for the optimized configuration in the on and the off states, are shown in fig. 7. Fig. 7(a)-7(d) shows that the structure has a good mode confinement in the structure moreover fig. 7(e) and fig. 7(f) indicate that the time-averaged Poynting's vector in the n-Si and the $BaTiO_3$ layers increase for the off state which it causes that the sensitivity of the modulator increases for changing the absorption coefficient in the active region.

## 4. Conclusion

In this study, an absorption plasmonic modulator is introduced and its operational parameters are numerically examined. It is noteworthy that for more realistic simulation, the Pockels effect at the $BaTiO_3$ and the free carrier dispersion effect at the n-type silicon are considered simultaneously. Here, the related parameters of the absorption modulation are investigated as a function of different thicknesses. The maximum value of the figure of merit, which occurs at an Air($\infty$)/Au(50nm)/$BaTiO_3$(30nm)/n-Si(12nm)/Au(50nm)/TSNAO($\infty$) multilayer structure, is equal to 12.79. Compared with the designed and fabricated plasmonic modulators [36,38,44,45,46], the FoM of our modulator is high which predicts higher performance with respect to them. By applying voltage, the energy flow increases in the n-Si and the barium titanate layers, hence, it causes that the plasmonic mode is more sensitive for changing the refractive indexes of these two layers. According to our results, the presented modulator is promising for exploiting as an intensity modulator, so it can be used in the plasmonic integrated circuits and due to compatibility with CMOS technology it can be integrated with microelectronic systems.

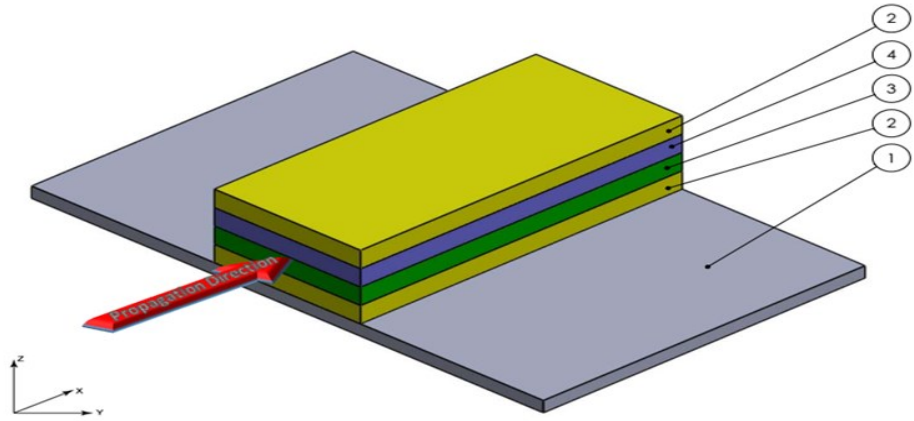

Figure 1. Schematic structure of designed modulator; (1) thin-shell nanolattice $Al_2O_3$ with near unity refractive index substrate (2) gold layer (3) n-Si layer (4) $BaTiO_3$ layer.

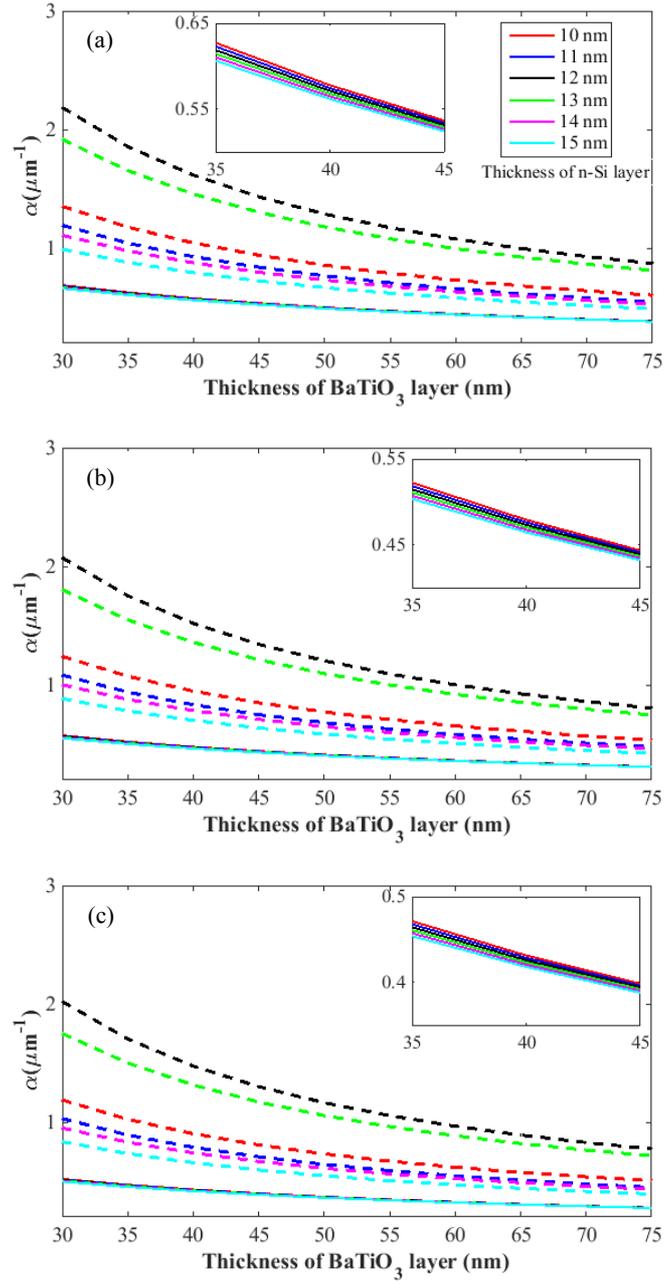

Fig. 2. Absorption coefficient of the structure for different thicknesses of the BaTiO$_3$ and the n-Si layers at 30nm (a), 40nm (b) and 50nm (c) thickness of gold layer for on (solid line) and off (dashed line) states. Each color indicates different thickness of the n-Si layer which specified in the fig. 2(a) as inset.

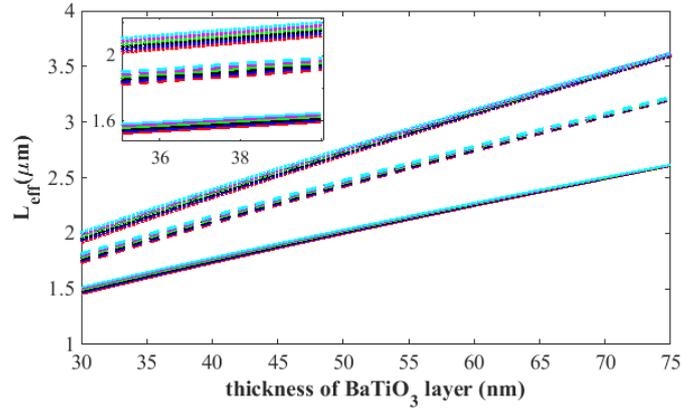

Fig. 3. Effective propagation length of the structure for different thicknesses of the BaTiO$_3$ and the n-Si layers at 30nm (solid line), 40nm (dashed line) and 50nm (dotted line) thickness of the gold layer. Colors represent the same thicknesses of the n-Si layers as fig. 2.

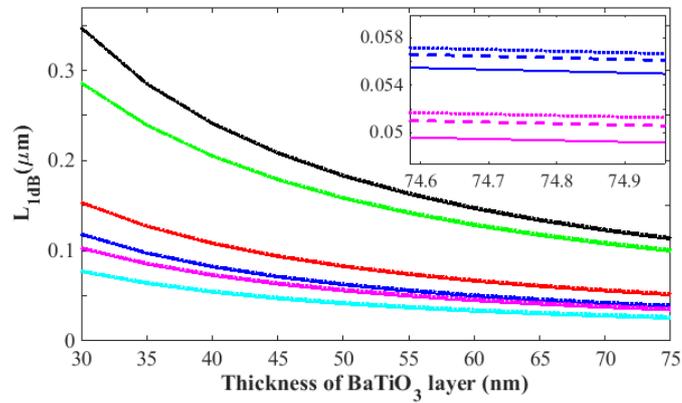

Fig. 4. 1dB on-off length of the structure is plotted for different thicknesses of the BaTiO$_3$ and n-Si layers at 30nm (solid line), 40nm (dashed line) and 50nm (dotted line) thickness of the gold layer. Colors represent the same thicknesses of the n-Si layers as fig. 2.

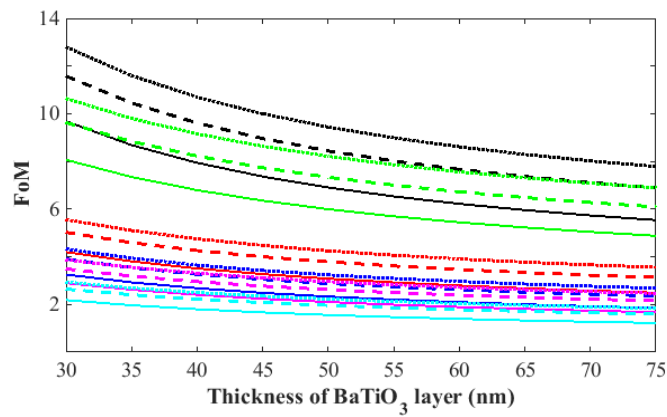

Fig. 5. Figure of merit (FoM) of the structure for different thicknesses of the BaTiO$_3$ and the n-Si layers at 30nm (solid line), 40nm (dashed line) and 50nm (dotted line) thickness of the gold layer. Colors represent the same thicknesses of the n-Si layers as fig. 2.

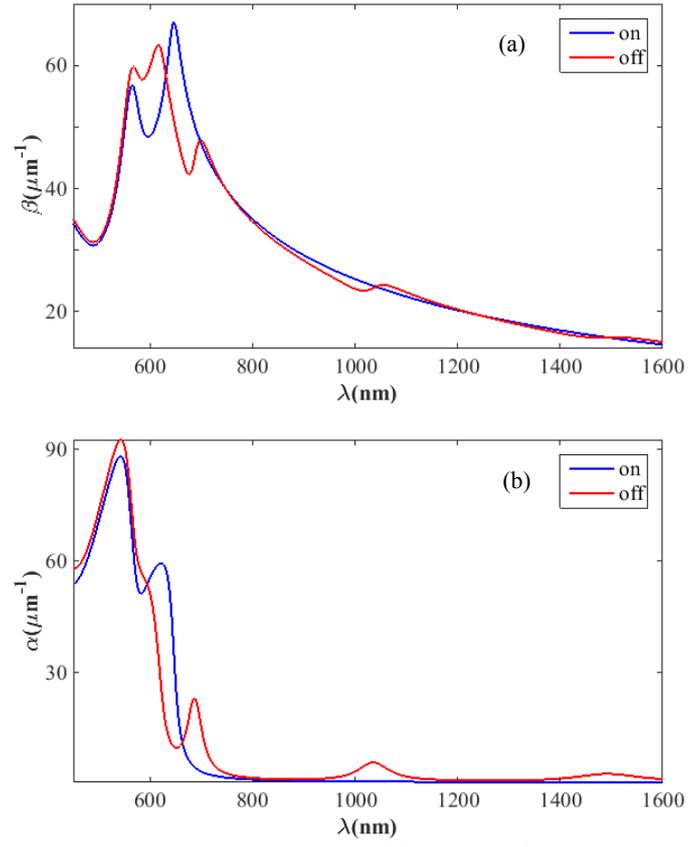

Fig 6. The propagation constant (a) and absorption coefficient (b) of the optimized modulator for the off (red line) and the on (blue line) states.

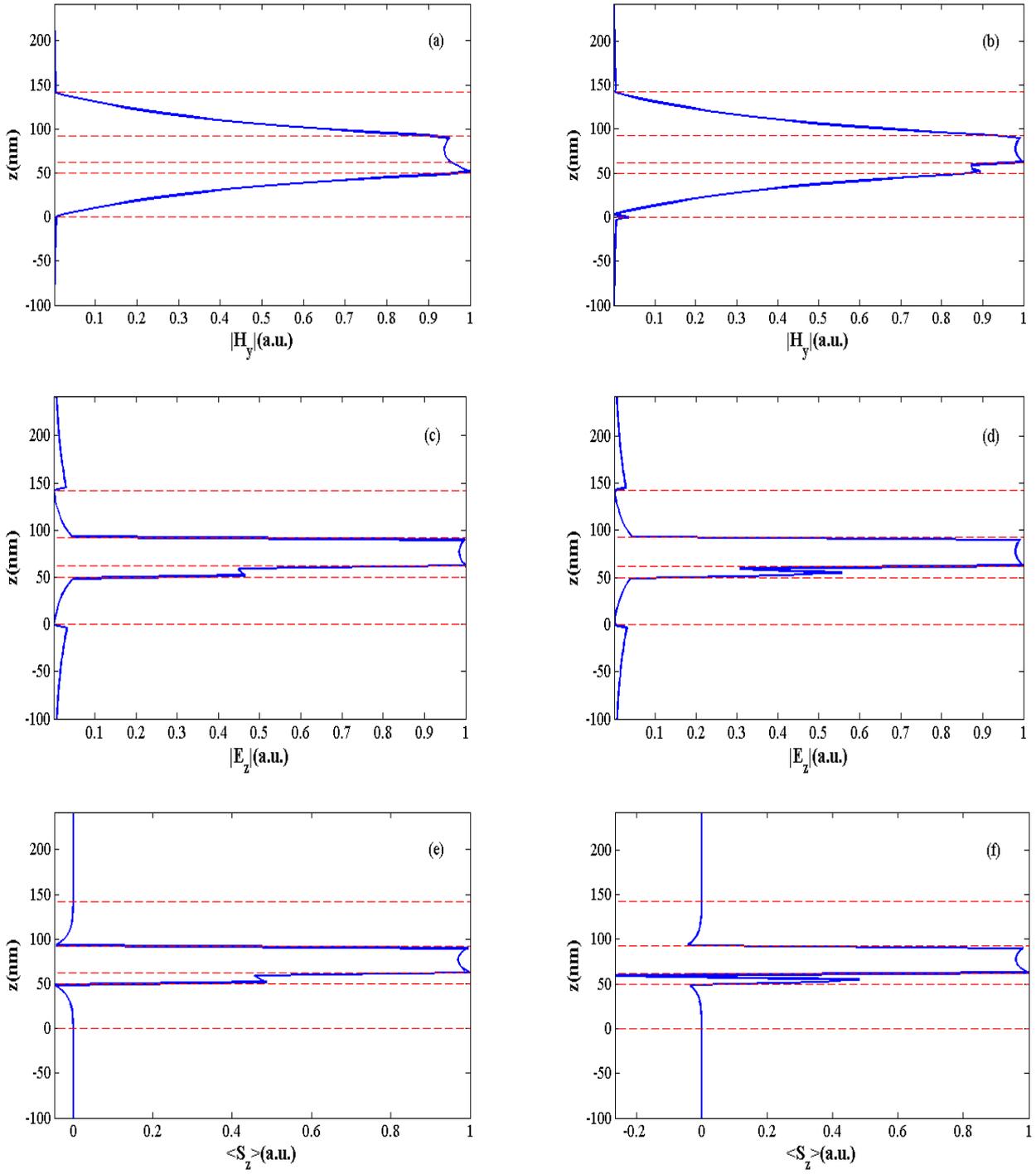

Fig. 7. The normalized magnetic field (a), z-component of the normalized electric field (c) and the normalized time-averaged Poynting's vector (e) of the optimized modulator are plotted for the on state at 1.55 micrometer while (b), (d) and (f) indicate the same parameters as (a), (c) and (e) but for the off state.

15